\documentclass{article}
\begin{document}
\setlength{\unitlength}{1mm}
\renewcommand{\baselinestretch}{1.0}
\newcommand{\Section}{\setcounter{equation}{0}\section}
\begin{center}
{\bf  Single-Particle Distributions in Multi-Chain Model with successive collision for High-Energy Nucleus-Nucleus collisions}

$\vspace{2cm}$

\footnotesize{$\rm{HUJIO\  NOD{A}^{\sharp,\P},\ SHIN-ICHI\ NAKARIK{I}^{+,\|}\ and\ TSUTOMU\ TASHIR{O}^{+,\dagger}}$

$\sharp\ Faculty\ of\ Science,\ Ibaraki\ University,\ Mito\ 310-8512,\ Japan$\footnote{Now, Emeritus Professor of Ibaraki University}

$+\ Research\ Institute\ of\ Natural\ Science,\ Okayama\ University\ of\ Science,\ Okayama\ $

$700-0005,\ Japan$


$\P:\ noda@mcs.ibaraki.ac,jp$

$\|:\ nakariki@sp.ous.ac.jp$

$\dagger:\ tashiro@sp.ous.ac.jp$
}

\end{center}

$\vspace{0.5cm}$

\small{High-energy nucleus-nucleus collisions are studied in multi-chain model with successive collision. Analytic forms for single-particle distribution are derived.}

$\vspace{0.5cm}$

\small{$Keywords$:$\ $Nucleus-nucleus collisions;$\ $Multi-chain model;$\ $ Single-particle distribution }

$\vspace{0.5cm}$

\small{\section{Introduction}}

Recently, nucleus-nucleus collisions at high energy have been studied very intensively because of the possibility of the formation of a quark-gluon plasma state. Also, the rapidity distributions of the charged particles and the ratios of the numbers of anti-hadrons to hadrons at SPS[1] and RHIC[2].

The nucleon-nucleus($N-A$)  and nucleus-nucleus($A-A$) collisions have been studied from the multiple-scattering view[3,4,5]. In particular, the multi-chain model(MCM) succeeded  phenomenologically to reproduce the old data on the multiplicity and the inclusive spectra of the leading nucleon and secondary particles  etc. in $ N-A$ and $A-A$ collisions[6,7,8,9]. In MCM[10], it is assumed that the nucleon-nucleon($N-N$) interaction is considered as a exchange of one multi peripheral chain from which secondary hadrons are emitted. However, it was difficult to perform analytical calculation for $A-A$ collisions because of the complexity of the general formula in resolving the full combination of chain configuration. Thus, dynamical simulation models for $A-A$ collisions have been developed.

In this paper, on the basis of MCM with successive collision[11], the full combinations of chain configuration are resolved in terms of the vector-operator method in the moment space by consulting the Mellin transformation[12]. The analytic forms for single-particle distributions in $A-A$ collisions are derived.   

In this paper, the dynamics is assumed to scale with energy.  A nucleus is treated as a set of mutually independent nucleons. The cascading of the produced hadrons is neglected owing to the long formation length in nucleus. Also, the longitudinal motion is treated. The transverse momentum distribution is assumed to be independent of  incident energy and longitudinal momentum.

In Section 2, the MCM with successive collision for $N-A$ collisions is summarized and the vector-operator formalism in the moment space is given.
In Section 3, single-particle distributions are investigated and their analytic forms are derived.
 In Section 4, conclusion and discussion are given.

\vspace{0.5cm}

\small{\section{MCM with successive collision for $N-A$ collisions}} 

\small{\subsection{Structure of the inelastic $N-N$ interaction}}

We start with the inelastic $N-N$ interaction. We assume that the inelastic $N-N$ interaction as the exchange of one chain with a mesonic cluster from which secondary hadrons are emitted for simplicity. Namely, we consider the reaction

\[N+N \to N+N+M \to N+N+ {\rm hadrons}\]
where $M$ denotes the mesonic cluster. We pay attention to the single-particle distribution of the inclusive process $ N+N \to C+X$(anything) given by
\begin{eqnarray}
\rho ^{NN}_C(x)=x\frac{d\sigma (NN \to CX)}{dx}\ \ \ \ \ \ \ \ ( x>0)
\end{eqnarray}
where $x$ denotes the longitudinal momentum fraction(Feynman variable). We express the projectile(target) fragmentation regions as $x>0\ (\ x<0\ )$.

By the Mellin transformation of Eq.(1), we define the moment $\rho^{NN}_C(J)$ and the operator $\hat {\rho}^{NN}(J)$ in the moment space as 
\begin{eqnarray}
\rho^{NN}_C (J)=\int_0^1 dxx^{J-2}\rho^{NN}_C (x)\equiv <C|\hat{\rho}^{NN}(J)|N>
\end{eqnarray}
where $|N>$ denotes the nucleon state in the projectile nucleus A. Here, we employ a vector-operator notation[12]. We introduce the operator $\hat{J}(J)$ for the inelastic interaction of $N-N$collisions as 
 \begin{eqnarray} 
\hat{J}(J) \equiv \frac{1}{\sigma^{NN}_{inel}}\hat{\rho}^{NN}(J)
\end{eqnarray} 
where $\sigma^{NN}_{inel}$ is the inelastic cross section of $N-N$ collisions.

We define the matrix elements of the leading nucleon $N$ which is not newly emitted from the mesonic cluster and the mesonic cluster $M$ as
\begin{eqnarray}
<N|\hat{J}(J)|N>=\frac{1}{\sigma^{NN}_{inel}}<N|\hat{\rho}^{NN}(J)|N> \equiv F(J),
\end{eqnarray}
\begin{eqnarray}
<M|\hat{J}(J)|N>=\frac{1}{\sigma^{NN}_{inel}}<M|\hat{\rho}^{NN}(J)|N> \equiv K(J) 
\end{eqnarray}
where $<N|$ and $<M|$ denote the leading nucleon state and the mesonic cluster state, respectively.
By the inverse Mellin transformation, the fraction functions $F(x)$ and $K(x)$ are given by
\begin{eqnarray}
F(x)=\frac{1}{2\pi i}\int_{c-i\infty}^{c+i\infty}dJx^{1-J}F(J),
\end{eqnarray}
\begin{eqnarray}
K(x)=\frac{1}{2\pi i}\int_{c-i\infty}^{c+i\infty}dJx^{1-J}K(J)
\end{eqnarray}
where  $ \int_{0}^{1} \frac{dx}{x}F(x)=1$ and $K(x)=\frac{x}{1-x}F(1-x)$ from the cascade mechanism.

\vspace{0.5cm}

\small{\subsection{MCM with successive collision and recurrence equations}} 

In MCM with successive collision for $N-A$ collisions, the basic assumption is that only the projectile nucleon can interact successively with the nucleons inside the target nucleus A, while nucleons inside nucleus A experience only one inelastic collision with projectile nucleon. This model can reproduce the single-particle distributions of the leading nucleon and secondary particles.
We show a sketch of MCM  in Fig.1.

First, according to Ref.[11], we summarize the MCM with successive collision for $N-A$ collisions at the impact parameter $\vec{b}$. The inclusive distributions of leading nucleon and mesonic cluster after $n$-times successive collisions of the projectile nucleon with nucleon inside the target nucleus A in the projectile fragmentation regions($x>0$) are given by $Q(x,n;\vec{b})$ and $M(x,n;\vec{b})$, respectively. They satisfy the integral equations
\begin{eqnarray} 
Q(x,n;\vec{b})=(1-\lambda(\vec{b})Q(x,n-1;\vec{b})+\lambda(\vec{b})\int_{x}^{1}\frac{dy}{y}F(\frac{x}{y})Q(y,n-1;\vec{b}),
\end{eqnarray} 
\begin{eqnarray}
M(x,n;\vec{b})=M(x,n-1;\vec{b})+\lambda(\vec{b})\int_{x}^{1}\frac{dy}{y}K(\frac{x}{y})Q(y,n-1;\vec{b})
\end{eqnarray}
where the functions $F(x/y)$ and $K(x/y)$ are given by Eqs.(6) and (7). Also, $\lambda(\vec{b})$ is the inelastic interaction probability defined as $\lambda(\vec{b})=\sigma^{NN}_{inel}T_A(\vec{b})$ where $T_A(\vec{b})$ is the nuclear thickness function of nucleus A normalized to unity. It is noted that the collision number $n$ is limited to $0\le n \le A$ where $A$ denotes the mass number of the nucleus A. 

By the iteration method, we obtain the following solutions of Eqs.(8) and (9) in the final collision($n=A$):

\begin{eqnarray}
Q(x,A;\vec{b})=P_{0}(A;\vec{b}) \delta (1-x)+\sum_{m=1}^{A}P_{m}(A;\vec{b}) F^{(m)}(x),
\end{eqnarray}
\begin{eqnarray}
M(x,A;\vec{b})=\sum_{m=1}^{A}P_{m}(A;\vec{b})\sum_{l=0}^{m-1} \int_{x}^{1} \frac{dy}{y} K(\frac{x}{y})F^{(l)}(y)
\end{eqnarray}

\begin{picture}(100,120)
\thicklines
\put(30,40){\line(60,0){60}}
\put(30,45){\line(60,0){60}}
\put(30,50){\line(60,0){60}}
\put(30,70){\line(60,0){60}}

\put(40,70){\line(-2,1){4}}
\put(40,70){\line(-2,-1){4}}
\put(80,70){\line(-2,1){4}}
\put(80,70){\line(-2,-1){4}}
\put(40,50){\line(-2,1){4}}
\put(40,50){\line(-2,-1){4}}
\put(40,45){\line(-2,1){4}}
\put(40,45){\line(-2,-1){4}}
\put(40,40){\line(-2,1){4}}
\put(40,40){\line(-2,-1){4}}
\put(80,50){\line(-2,1){4}}
\put(80,50){\line(-2,-1){4}}
\put(80,45){\line(-2,1){4}}
\put(80,45){\line(-2,-1){4}}
\put(80,40){\line(-2,1){4}}
\put(80,40){\line(-2,-1){4}}

\put(50,50){\circle*{1.4}}
\put(50,70){\circle*{1.4}}
\put(65,45){\circle*{1.4}}
\put(65,70){\circle*{1.4}}

\put(50,51){\oval(2,2)[r]}
\put(50,53){\oval(2,2)[l]}
\put(50,55){\oval(2,2)[r]}
\put(50,57){\oval(2,2)[l]}
\put(50,59){\oval(2,2)[r]}
\put(50,61){\oval(2,2)[l]}
\put(50,63){\oval(2,2)[r]}
\put(50,65){\oval(2,2)[l]}
\put(50,67){\oval(2,2)[r]}
\put(50,69){\oval(2,2)[l]}

\put(65,46){\oval(2,2)[r]}
\put(65,48){\oval(2,2)[l]}
\put(65,50){\oval(2,2)[r]}
\put(65,52){\oval(2,2)[l]}
\put(65,54){\oval(2,2)[r]}
\put(65,56){\oval(2,2)[l]}
\put(65,58){\oval(2,2)[r]}
\put(65,60){\oval(2,2)[l]}
\put(65,62){\oval(2,2)[r]}
\put(65,64){\oval(2,2)[l]}
\put(65,66){\oval(2,2)[r]}
\put(65,68){\oval(2,2.5)[l]}

\put(25,70){N}
\put(25,45){A}
\put(90,15){Fig.1}

\end{picture}

 Fig.1 MCM with one kind of chain for $N-A$ collisions.  Wavy lines represent the inelastic $N-N$ interaction.

\vspace{0.5cm}

where $F^{(m)}(x)=\int_{x}^{1} \frac{dy}{y}F(\frac{x}{y})F^{(m-1)}(y)$ and $F^{(0)}(y)=\delta(1-y)$.
Also, $P_{m}(A;\vec{b})=\left (\matrix{A \cr
           m \cr} \right )(1-\lambda(\vec{b}))^{A-m} \lambda(\vec{b})^m $ which is the Glauber probability. The inelastic cross section of $N-A$ collisions is given by $\sigma^{NA}_{inel}=\int d\vec{b}\sum_{m=1}^{A} P_{m}(A;\vec{b})$ and the averaged collision number $\bar{n}=A\sigma^{NN}_{inel}/\sigma^{NA}_{inel}$. The moments of Eqs.(10) and (11) are given by
\begin{eqnarray}
Q(J,A;\vec{b})=P_0(A;\vec{b})+\sum^{A}_{m=1} P_m(A;\vec{b})F(J)^m,
\end{eqnarray}
\begin{eqnarray}
M(J,A;\vec{b})=\sum^{A}_{m=1} P_m(A;\vec{b})\sum_{l=0}^{m-1}K(J)F(J)^l.
\end{eqnarray}

Next, we reformulate the above MCM with successive collision in order to extend the model to $A-A$ collisions.

(i) The operator $\hat{Q}_A(J,n;\vec{b})$ is introduced and the moments $Q(J,n;\vec{b})$ and $M(J,n;\vec{b})$ are defined as follows: 
\begin{eqnarray}
Q(J,n;\vec{b}) \equiv <N|\hat{Q}_A(J,n;\vec{b})|N> ,
\end{eqnarray}
\begin{eqnarray}
M(J,n;\vec{b}) \equiv <M|\hat{Q}_A(J,n;\vec{b})|N>.
\end{eqnarray}

We assume the following recurrence equations for $\hat{Q}(J,n;\vec{b})$ on the analogy with Eq.(8) in the moment space: 
\begin{eqnarray}
\hat{Q}_A(J,n;\vec{b})=[(1-\lambda(\vec{b}))+\lambda(\vec{b})\hat{J}(J)]\hat{Q}_A(J,n-1;\vec{b})
\end{eqnarray}
where $\hat{Q}(J,n;\vec{b})=1$. From Eq.(16), we get the solution
\begin{eqnarray}
\hat{Q}_A(J,n;\vec{b})=[(1-\lambda(\vec{b}))+\lambda(\vec{b})\hat{J}(J)]^n.
\end{eqnarray}
Then, we put $n=A$ and use the binomial expansion. Eq.(17) reduces to
\begin{eqnarray}
\hat{Q}_A(J,A;\vec{b})=P_0(A;\vec{b})+\sum_{m=1}^{A}P_{m}(A;\vec{b})\hat{J}(J)^m.
\end{eqnarray}
where $m$ agrees with the number of chain in MCM.

(ii) We define the matrix elements for $\hat{J}(J)^m$ in Eq.(18) as 
\begin{eqnarray}
N(J,m)\equiv <N|\hat{J}(J)^m|N>,
\end{eqnarray}
\begin{eqnarray}
M(J,m)\equiv <M|\hat{J}(J)^m|N>
\end{eqnarray}
where $N(J,0)=1$ and $M(J,0)=0$. Also, $N(J,1)=F(J)$ and $M(J,1)=K(J)$ from Eqs.(4) and (5).

We assume the recurrence equations
\begin{eqnarray}
N(J,m)=F(J)N(J,m-1),
\end{eqnarray}
\begin{eqnarray}
M(J,m)=M(J,m-1)+K(J)N(J,m-1).
\end{eqnarray}
Eq.(21) satisfies the baryon number conservation and Eq.(22) is derived on the analogy with Eq.(9) in the moment space. The solutions of Eqs.(21) and (22) are given by
\begin{eqnarray}
N(J,m)=F(J)^m,
\end{eqnarray}
\begin{eqnarray}
M(J,m)=\sum_{l=0}^{m-1}K(J)F(J)^l.
\end{eqnarray}
Therefore, from Eqs.(18) and (23), Eq.(12) is derived and from Eqs.(18) and (24), Eq.(13) is derived. Thus, Eqs.(10) and (11) of MCM with successive collision for $N-A$ collisions are derived from the three recurrence equations of Eqs.(16), (21) and (22).
\vspace{0.5cm}

\small{\section{Single-particle distributions in $A-A$ collisions}} 

We investigate single-particle distributions in $A-A$ collisions on the basis of Eqs.(16), (21) and (22) for $N-A$ collisions. Consider the collision of two nuclei with mass numbers $A$ and $B$($A-B$ collisions) at the impact parameter $\vec{b}$. We treat the inclusive processes of $AB \to NX$ and $MX$ for the projectile fragmentation regions of nucleus A($x>0$).

First, we introduce the operator  $\hat{Q}_{AB}(J;\vec{b},A)$ in the moment space. Consider the moments $Q^{AB}_{N/N}(J;\vec{b},A)$, $M^{AB}_{M/N}(J;\vec{b},A)$ of the inclusive distributions $Q^{AB}_{N/N}(x;\vec{b},A)$ and $M^{AB}_{M/N}(x;\vec{b},A)$. They are defined as 
\begin{eqnarray}
Q^{AB}_{N/N}(J;\vec{b},A)=<N|\hat{Q}_{AB}(J;\vec{b},A)|N>,
\end{eqnarray}
\begin{eqnarray}
M^{AB}_{M/N}(J;\vec{b},A)=<M|\hat{Q}_{AB}(J;\vec{b},A)|N>
\end{eqnarray}
where the argument $A$ denotes the treatment of  particles observed in the fragmentation regions of nucleus A in final state. The $|N>$ is the nucleon state inside the projectile nucleus A and the states $<N|$ and $<M|$ denote the leading nucleon and the mesonic cluster observed in the projectile fragmentation regions of nucleus A in the initial state, respectively. 

Second, we introduce the operator $\hat{Q}_{AB}(J,n;\vec{b})$ which characterizes the total structure of $A-A$ collisions. The recurrence equation is assumed as follows
\begin{eqnarray}
\hat{Q}_{AB}(J,n;\vec{b})=[(1-\lambda(\vec{b}))\hat{G(J)}+\lambda(\vec{b})\hat{J}(J)]\hat{Q}_{AB}(J,n-1;\vec{b}),
\end{eqnarray}
where  $n$ is the collision number$(0 \le n \le AB)$ and $\hat{Q}_{AB}(J,0;\vec{b})=1$. Also, $\lambda(\vec{b})=\sigma^{NN}_{inel}\int d\vec{s} \int d\vec{t} T_A(\vec{s})T_B(\vec{t})\delta^2(\vec{b}+\vec{s}-\vec{t})$ where $T_A(\vec{s})$ and $T_{B}(\vec{t})$ are the normalized nuclear thickness functions of nucleus A and nucleus B, respectively. We introduce the operator $\hat{G}(J)$ so as to express the passing-through probability $(1-\lambda(\vec{b}))$ explicitly. By considering both operators $\hat{G}(J)$ and $\hat{J}(J)$, we may calculate full combinations of the chain configurations with ease. In fact,  it corresponds to MCM with two kinds of chain as shown in Fig.2.

The solution of Eq.(27) is given by
\begin{eqnarray}
\hat{Q}_{AB}(J,n;\vec{b})=[(1-\lambda(\vec{b}))\hat{G(J)}+\lambda(\vec{b})\hat{J}(J)]^{n}.
\end{eqnarray}
The operator $\hat{Q}_{AB}(J;\vec{b})$ is defined to the limit
\begin{eqnarray}
\hat{Q}_{AB}(J;\vec{b})=\lim_{n \to AB}\hat{Q}_{AB}(J,n;\vec{b}).
\end{eqnarray}

Taking notice of the ordering of $\hat{G}(J)$ and $\hat{J}(J)$, we expand $\hat{Q}_{AB}(J;\vec{b})$. Then, Eq.(29) reduces to
\[\hat{Q}_{AB}(J;\vec{b})=(1-\lambda(\vec{b}))^{AB}\hat{G}(J)^{AB}+(1-\lambda(\vec{b}))^{AB-1}\lambda(\vec{b})[\hat{G}(J)^{AB-1}\hat{J}(J)\]
\begin{eqnarray}
+\hat{G}(J)^{AB-2}\hat{J}(J)\hat{G}(J)+ \cdots +\hat{J}(J)\hat{G}(J)^{AB-1}]+ \cdots +\lambda(\vec{b})^{AB}\hat{J}(J)^{AB}.
\end{eqnarray}

\begin{picture}(120,120)
\thicklines
\put(30,40){\line(60,0){60}}
\put(30,45){\line(60,0){60}}
\put(30,50){\line(60,0){60}}
\put(30,70){\line(60,0){60}}
\put(30,75){\line(60,0){60}}

\put(40,75){\line(-2,1){4}}
\put(40,75){\line(-2,-1){4}}
\put(80,75){\line(-2,1){4}}
\put(80,75){\line(-2,-1){4}}

\put(40,70){\line(-2,1){4}}
\put(40,70){\line(-2,-1){4}}
\put(80,70){\line(-2,1){4}}
\put(80,70){\line(-2,-1){4}}

\put(40,50){\line(-2,1){4}}
\put(40,50){\line(-2,-1){4}}
\put(40,45){\line(-2,1){4}}
\put(40,45){\line(-2,-1){4}}
\put(40,40){\line(-2,1){4}}
\put(40,40){\line(-2,-1){4}}

\put(80,50){\line(-2,1){4}}
\put(80,50){\line(-2,-1){4}}
\put(80,45){\line(-2,1){4}}
\put(80,45){\line(-2,-1){4}}
\put(80,40){\line(-2,1){4}}
\put(80,40){\line(-2,-1){4}}

\put(45,45){\circle*{1.4}}
\put(45,75){\circle*{1.4}}
\put(50,50){\circle*{1.4}}
\put(50,70){\circle*{1.4}}
\put(55,40){\circle*{1.4}}
\put(55,75){\circle*{1.4}}
\put(60,40){\circle*{1.4}}
\put(60,70){\circle*{1.4}}
\put(65,70){\circle*{1.4}}
\put(65,45){\circle*{1.4}}
\put(70,75){\circle*{1.4}}
\put(70,50){\circle*{1.4}}

\put(50,51){\oval(2,2)[r]}
\put(50,53){\oval(2,2)[l]}
\put(50,55){\oval(2,2)[r]}
\put(50,57){\oval(2,2)[l]}
\put(50,59){\oval(2,2)[r]}
\put(50,61){\oval(2,2)[l]}
\put(50,63){\oval(2,2)[r]}
\put(50,65){\oval(2,2)[l]}
\put(50,67){\oval(2,2)[r]}
\put(50,69){\oval(2,2)[l]}

\put(65,46){\oval(2,2)[r]}
\put(65,48){\oval(2,2)[l]}
\put(65,50){\oval(2,2)[r]}
\put(65,52){\oval(2,2)[l]}
\put(65,54){\oval(2,2)[r]}
\put(65,56){\oval(2,2)[l]}
\put(65,58){\oval(2,2)[r]}
\put(65,60){\oval(2,2)[l]}
\put(65,62){\oval(2,2)[r]}
\put(65,64){\oval(2,2)[l]}
\put(65,66){\oval(2,2)[r]}
\put(65,68){\oval(2,2.5)[l]}

\put(55,42){\oval(2,2)[r]}
\put(55,44){\oval(2,2)[l]}
\put(55,46){\oval(2,2)[r]}
\put(55,48){\oval(2,2)[l]}
\put(55,50){\oval(2,2)[r]}
\put(55,52){\oval(2,2)[l]}
\put(55,54){\oval(2,2)[r]}
\put(55,56){\oval(2,2)[l]}
\put(55,58){\oval(2,2)[r]}
\put(55,60){\oval(2,2)[l]}
\put(55,62){\oval(2,2)[r]}
\put(55,64){\oval(2,2)[l]}
\put(55,66){\oval(2,2)[r]}
\put(55,68){\oval(2,2)[l]}
\put(55,70){\oval(2,2)[r]}
\put(55,73){\oval(2,3)[l]}

\put(45,45){\line(0,2){2}}
\put(45,48){\line(0,2){2}}
\put(45,51){\line(0,2){2}}
\put(45,54){\line(0,2){2}}
\put(45,57){\line(0,2){2}}
\put(45,60){\line(0,2){2}}
\put(45,63){\line(0,2){2}}
\put(45,66){\line(0,2){2}}
\put(45,69){\line(0,2){2}}
\put(45,71){\line(0,2){2}}
\put(45,74){\line(0,2){2}}

\put(60,40){\line(0,2){2}}
\put(60,43){\line(0,2){2}}
\put(60,46){\line(0,2){2}}
\put(60,49){\line(0,2){2}}
\put(60,52){\line(0,2){2}}
\put(60,55){\line(0,2){2}}
\put(60,58){\line(0,2){2}}
\put(60,61){\line(0,2){2}}
\put(60,64){\line(0,2){2}}
\put(60,67){\line(0,2){2}}

\put(70,50){\line(0,2){2}}
\put(70,53){\line(0,2){2}}
\put(70,54){\line(0,2){2}}
\put(70,58){\line(0,2){2}}
\put(70,61){\line(0,2){2}}
\put(70,64){\line(0,2){2}}
\put(70,67){\line(0,2){2}}
\put(70,70){\line(0,2){2}}
\put(70,73){\line(0,2){2}}
\put(25,72.5){A}
\put(25,45){B}
\put(90,15){Fig.2}
\end{picture}

 Fig.2 MCM with two kinds of chain for $A-B$ collisions.  Wavy lines and dashed lines represent the inelastic interaction and the passing-through in $N-N$ collision, respectively.

\vspace{0.5cm}

If $\hat{G}(J)=1,$ Eq.(30) is given by
\begin{eqnarray}
\hat{Q}(J;\vec{b})=P_{0}(AB;\vec{b})+\sum_{m=1}^{AB}P_{m}(AB;\vec{b}) \hat{J}(J)^{m}
\end{eqnarray}
where $ P_m(AB;\vec{b})=\left(\matrix{
                                                                AB \cr
                                                                m \cr} \right)(1-\lambda(\vec{b}))^{AB-m}\lambda(\vec{b})^m$
that is the Glauber probability for $A-B$ collisions and $m$ agrees with the number of chain in MCM. The inelastic cross section of $A-B$ collisions is given by $\sigma^{NN}_{inel}=\int d\vec{b}\sum_{m=1}^{AB} P_{m}(AB:\vec{b})$ and the averaged collision number $\bar{n}=AB\sigma^{NN}_{inel}/\sigma^{AB}_{inel}$. 

Third, we discuss the relationship between $\hat{Q}_{AB}(J;\vec{b},A)$ and $\hat{Q}_{AB}(J;\vec{b})$. In MCM, the maximum number of collision of one nucleon inside the projectile nucleus A is restricted to the mass number of the target nucleus B as investigated in $N-A$ collisions  and vice versa.
Thus, we must impose the restriction to $\hat{Q}_{AB}(J;\vec{b})$. When we observe the particle in the projectile fragmentation regions, we introduce the rule of the sum decomposition in the projectile fragmentation regions of nucleus A for
the operator product $ \hat{O}_{1}\hat{O}_{2} \cdots \hat{O}_{AB} $  in Eq.(30) as follows:
\begin{eqnarray}
:\hat{O}_{1}\hat{O}_{2} \cdots \hat{O}_{AB}:_{A}=\sum_{i=1}^{A}\Pi_{j=1}^{B}\hat{O}_{j+(i-1)B}.
\end{eqnarray}
>From Eq.(32), we get
\[:\hat{Q}(J;\vec{b}):_{A}=(1-\lambda(\vec{b}))^{AB}A\hat{G}(J)^{B}+(1-\lambda(\vec{b}))^{AB-1}\lambda(\vec{b})[(A-1)\hat{G}(J)^{B}+\hat{G}(J)^{B-1}\hat{J}(J)\]
\begin{eqnarray}
+ \cdots +\hat{J}(J)\hat{G}(J)^{B-1}+(A-1)\hat{G}(J)^{B}]+\cdots +\lambda(\vec{b})^{AB}A\hat{J}(J)^{B}
\end{eqnarray}
Furthermore, we use the relation $[\hat{G}(J),\hat{J}(J)]=0$. Eq.(33) reduces to
\begin{eqnarray}
:\hat{Q}(J;\vec{b}):_{A}=A\sum_{m=0}^{AB}P_{m}(AB;\vec{b})\sum_{l=0}^{B}H(l;AB,m,B)\hat{G}(J)^{B-l}\hat{J}(J)^l
\end{eqnarray}
where $H(l;AB,m,B)$ is the hypergeometric distribution  defined as
\[
H(l;AB,m,B)=\frac{\left(\matrix{ AB-m \cr
 B-l \cr}\right) 
 \left(\matrix{m \cr
 l \cr }\right)}
 {\left(\matrix{ AB \cr
 B \cr}\right)}.
\]
The operator $\hat{Q}_{AB}(J;\vec{b},A)$ is defined as 
\begin{eqnarray}
\hat{Q}_{AB}(J;\vec{b},A)=:\hat{Q}(J;\vec{b}):_{A}.
\end{eqnarray}
Thus, from Eq.(34), we obtain
\begin{eqnarray}
\hat{Q}_{AB}(J;\vec{b},A)=A\sum_{m=0}^{AB}P_{m}(AB;\vec{b})\sum_{l=0}^{B}H(l;AB,m,B)\hat{J}(J)^l
\end{eqnarray}
where we put $\hat{G}(J)=1$. From Eqs.(25), (26) and (36),
\begin{eqnarray}
Q^{AB}_{N/N}(J;\vec{b},A)=A\sum_{m=0}^{AB}P_{m}(AB;\vec{b})\sum_{l=0}^{B}H(l;AB,m,B)F(J)^l,
\end{eqnarray}
\begin{eqnarray}
M^{AB}_{M/N}(J;\vec{b},A)=A\sum_{m=1}^{AB}P_{m}(AB;\vec{b})\sum_{l=1}^{B}H(l;AB,m,B)\sum_{r=0}^{l-1}K(J)F(J)^r.
\end{eqnarray}
By means of the inverse Mellin transformation of Eqs.(37) and (38), the single-particle distributions of $AB \to NX$ and $MX$ for the projectile fragmentation regions($x>0$) are given by
\begin{eqnarray}
\rho^{AB}_{N}(x)=\int d\vec{b}A\sum_{m=1}^{AB}P_{m}(AB:\vec{b})\sum_{l=1}^{B}H(l;AB,m,B)F^{(l)}(x),
\end{eqnarray}
\begin{eqnarray}
\rho^{AB}_M(x)=\int d\vec{b}A\sum_{m=1}^{AB}P_{m}(AB:\vec{b})\sum_{l=1}^{B}H(l;AB,m,B)\sum_{r=1}^{l-1}\int_x^1\frac{dy}{y}K(\frac{x}{y})F^{(r)}(x).
\end{eqnarray}
We obtain the similar results for the target fragmentation regions($x<0$). It is easy to show that Eqs.(37) and (38) reduce to Eqs.(12) and (13) for $x>0$
in $N-A$ collisions. Also, we obtain 
\begin{eqnarray}
\rho^{NA}_{N}(x)=A\sigma^{NN}_{inel}F(x),
\end{eqnarray}
\begin{eqnarray}
\rho^{NA}_{M}(x)=A\sigma^{NN}_{inel}K(x)
\end{eqnarray}
for the target fragmentation regions of nucleus A$(x<0 )$.

\vspace{0.5cm}

\small{\section{Conclusion and Discussion }} 

We have investigated the single-particle distribution in $A-A$ collisions on the basis of MCM with successive collision. By assumption of three recurrence equations with the rule of the sum decomposition, we derive the analytic form of the single-particle distribution of the inclusive process $A+B \to C+X$ which is expressed in the Glauber probability with the hypergeometric distribution. 

This result may be interpreted from the probabilistic view as follows:
The total number of collisions in $A-B$ collisions is $AB$. The maximum number of collisions one nucleon inside projectile nucleon A is $B$. The number of the inelastic $N-N$ interaction( $\hat{J}(J)$ ) and the number of the passing-through ($\hat{G}(J)$) correspond to $m$ and $AB-m$, respectively. If the number of the inelastic $N-N$ interaction actually occurring in the event is $l$, the probability of occurring exactly $l$ can be calculated by the hypergeometric distribution($H(l;AB,m,B)$) .

Recently, the polarization phenomena in $A-A$ collisions are payed attention to and the global polarization[13]  is investigated. This theme is one of the interesting problems in future.

\end{document}